# A Band-independent Variable Step Size Proportionate Normalized Subband Adaptive Filter Algorithm

Yi Yu[1, 2], and Haiquan Zhao[1, 2]

**Abstract**

Proportionate-type normalized suband adaptive filter (PNSAF-type) algorithms are very attractive choices for echo cancellation. To further obtain both fast convergence rate and low steady-state error, in this paper, a variable step size (VSS) version of the presented improved PNSAF (IPNSAF) algorithm is proposed by minimizing the square of the noise-free *a posterior* subband error signals. A noniterative shrinkage method is used to recover the noise-free *a priori* subband error signals from the noisy subband error signals. Significantly, the proposed VSS strategy can be applied to any other PNSAF-type algorithm, since it is independent of the proportionate principles. Simulation results in the context of acoustic echo cancellation have demonstrated the effectiveness of the proposed method.

## 1 Introduction

Over the past few decades, adaptive filtering algorithms have received great deal of development and been widely applied in practical fields such as system identification, channel equalization, echo cancellation and beamforming [1], [2]. One of the popular algorithms is the normalized least mean square (NLMS) algorithm, due to its simplicity and easy implementation. However, it suffers from slow convergence when the input signal is colored,

[1] Key Laboratory of Magnetic Suspension Technology and Maglev Vehicle, Ministry of Education, Southwest Jiaotong University, Chengdu, 610031, China.

[2] School of Electrical Engineering at Southwest Jiaotong University, Chengdu, 610031, China.

E-mail: yuyi_xyuan@163.com, hqzhao@home.swjtu.edu.cn

especially for speech signal in echo cancellation. Aiming to the colored input signal, affine projection algorithm (APA) and some of its variants can speed up the convergence (e.g., see [3], [4], [27], [28] and the references therein) by utilizing the previous input vectors to update the tap-weight vector. Nevertheless, they require large computational cost due mainly to involving the matrix inversion operation. Another attractive approach is to use the subband adaptive filter (SAF) to deal with the colored input signal [5]. In SAF, the colored input signal is divided into almost mutually exclusive subband signals, and each subband signal is approximately white, thus improving the convergence. In [6], Lee and Gan proposed the normalized SAF (NSAF) algorithm based on the principle of the minimum perturbation, which has faster convergence rate than the NLMS for the colored input signal. Furthermore, for applications of long filter such as echo cancellation, the NSAF algorithm has almost the same computational complexity as the NLMS. In fact, the NSAF will be reduced to the NLMS when number of subbands is equal to one. In [7], Yin *et al.* studied the convergence models of the NSAF in the mean and mean-square senses by assuming that the analysis filter bank is paraunitary and using several hyperelliptic integrals. On another hand, to overcome the trade-off problem of the NSAF between the convergence rate and steady-state error, several variable step size (VSS) NSAF algorithms were presented [8]-[11], [24].

In many realistic applications, sparse systems are often encountered (e.g., the impulse response of the echo paths), which have the property that only a fraction coefficients of impulse response (called *active coefficients*) have large magnitude while the rest coefficients (called *inactive coefficients*) are zero or very small. To improve the convergence rate of the classic adaptive filtering algorithms in sparse systems, several proportionate-type algorithms were developed [12]-[17]. The fundamental principle of this kind of algorithms is that each coefficient of the adaptive filter receives an independent step size in proportion to its estimated magnitude. The first proportionate algorithm is the proportionate NLMS (PNLMS) proposed by Duttweiler [12], which obtains faster initial convergence rate than the NLMS for a sparse case. However, the PNLMS shows a slow convergence when the unknown impulse response is dispersive. Moreover, its fast initial convergence is not maintained over the whole adaptation process. To solve the first problem of the PNLMS, the improved PNLMS (IPNLMS) algorithm was proposed by combining the proportionate (PNLMS) adaptation with non-proportionate (NLMS) adaptation [13]. To maintain fast initial convergence rate during the whole estimation process, the $\mu$-law PNLMS (MPNLMS) algorithm was proposed in [14] by deriving the optimal step-size control rule. Recently, in order to enhance the convergence speed of the NSAF for sparse systems, Abadi *et al*. developed a class of proportionate NSAF algorithms by directly extending the existing proportionate ideas in the NLMS to the NSAF, e.g., the proportionate NASF (PNSAF), $\mu$-law PNSAF (MPNSAF), improved PNSAF (IPNSAF), and so forth [16], [17]. However, similar to the NSAF and/or the NLMS, the overall performance (including the convergence rate, tracking capability and steady-state error) of these PNSAF-type algorithms is dominated by a fixed step-size, i.e., a large step size results in faster convergence and tracking, while the steady-state error is reduced for a small step size. To address this problem, the set-membership IPNSAF (SM-IPNSAF) algorithm was proposed in [17], because it can be interpreted as a VSS algorithm. Inadequately, its convergence performance is sensitive to the selection of the error bound, and its improvement in the steady-state error is slight as compared to the IPNSAF.

To obtain both fast convergence rate and low steady-state error, this paper develops a VSS version of the IPNSAF. In this VSS algorithm, the individual time-varying step size for each subband is derived by minimizing the square of the noise-free *a posterior* subband error signal. Furthermore, the noise-free *a priori* subband error signal is obtained by using a noniterative shrinkage method reported in [18], [19]. More importantly, the proposed VSS scheme can be applied to other existing PNSAF-type algorithms to improve their performance, due to the fact that it does not depend on the proportionate rules. Besides, the convergence condition of PNSAF-type algorithms in mean-square sense is provided in Appendix.

## 2 PNSAF-type algorithms

Consider the desired signal $d(n)$ that arises from the unknown system,

$$d(n) = \mathbf{u}^T(n)\mathbf{w}_o + \eta(n), \tag{1}$$

where $(\cdot)^T$ indicates transpose of a vector or a matrix, $\mathbf{w}_o$ is the unknown $M$-dimensional vector to be identified with an adaptive filter, $\mathbf{u}(n) = [u(n), u(n-1), ..., u(n-M+1)]^T$ is the input signal vector, and $\eta(n)$ is the system noise with zero-mean and variance $\sigma_\eta^2$. Fig. 1 shows the block diagram of multiband-structured SAF, where $N$ denotes number of subbands. The desired signal $d(n)$ and the input signal $u(n)$ are partitioned into multiple subband signals $d_i(n)$ and $u_i(n)$ through the analysis filter bank $\{H_i(z), i \in [0, N-1]\}$, respectively. The subband output signals $y_i(n)$ are obtained from $u_i(n)$ filtered by the adaptive filter given by $\mathbf{w}(k) = [w_1(k), w_2(k), ..., w_M(k)]^T$. Then, $y_{i,D}(k)$ and $d_{i,D}(k)$ are generated by critically decimating $y_i(n)$ and $d_i(n)$. Here, $n$ and $k$ are used to indicate the original sequences and the decimated sequences, respectively. It is easy to note that $y_{i,D}(k) = \mathbf{u}_i^T(k)\mathbf{w}(k)$, where $\mathbf{u}_i(k) = [u_i(kN), u_i(kN-1), ..., u_i(kN-M+1)]^T$. Accordingly, the $i$th subband error signal is given by

$$e_{i,D}(k) = d_{i,D}(k) - y_{i,D}(k) = d_{i,D}(k) - \mathbf{u}_i^T(k)\mathbf{w}(k) \tag{2}$$

where $d_{i,D}(k) = d_i(kN)$.

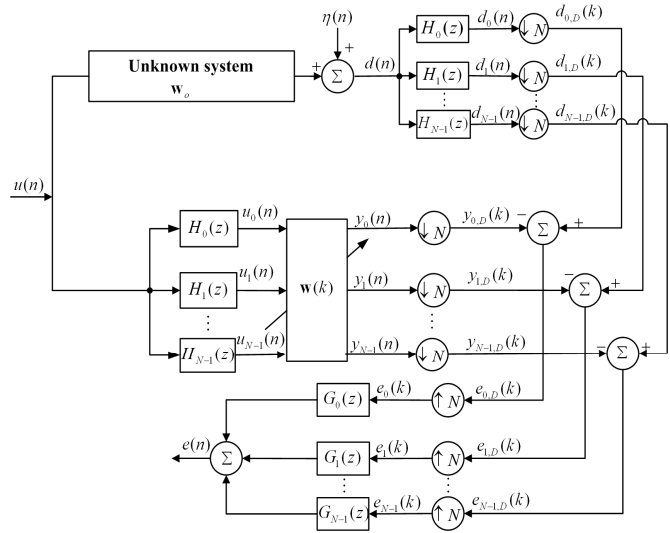

**Fig. 1.** Block diagram of multiband-structured SAF

For all the PNSAF-type algorithms [16], [17], the update formula of the tap-weight vector is expressed as

$$\mathbf{w}(k+1) = \mathbf{w}(k) + \mu \sum_{i=0}^{N-1} \frac{\mathbf{G}(k)\mathbf{u}_i(k)e_{i,D}(k)}{\mathbf{u}_i^T(k)\mathbf{G}(k)\mathbf{u}_i(k) + \delta} \tag{3}$$

where $\mu$ is the step-size, $\delta$ is the regularization constant to avoid division by zero, and $\mathbf{G}(k) = \text{diag}\{g_1(k), g_2(k), ..., g_M(k)\}$ is an $M \times M$ diagonal matrix (called the proportionate matrix) whose role is to assign an individual step size for each filter coefficient (i.e., a filter coefficient $w_m(k)$ with larger magnitude receives a larger step size $\mu g_m(k)$, thus improving the convergence rate of that coefficient). Evidently, different strategies to calculate the proportionate matrix $\mathbf{G}(k)$ can generate different PNSAF algorithms, e.g., [17]. In particular, the NSAF algorithm is obtained when $\mathbf{G}(k) = \mathbf{I}_{M \times M}$ with $\mathbf{I}_{M \times M}$ being the identity matrix (i.e., when all the filter coefficients receive the same increment).

## 3 Proposed VSS-IPNSAF

Now, we start to derive a VSS scheme which is suitable for any PNSAF-type algorithm, whose inspiration arises from the presented shrinkage NLMS (SHNLMS) algorithm in [18].

### 3.1 Derivation of VSS scheme

Replacing the fixed step size $\mu$ with an individual time-varying step size $\mu_i(k)$ for each subband and neglecting $\delta$ for the convenience of derivation, (3) becomes

$$\mathbf{w}(k+1) = \mathbf{w}(k) + \sum_{i=0}^{N-1} \mu_i(k) \frac{\mathbf{G}(k)\mathbf{u}_i(k)e_{i,D}(k)}{\mathbf{u}_i^T(k)\mathbf{G}(k)\mathbf{u}_i(k)}. \tag{4}$$

Before deriving this VSS scheme, we define the noise-free *a priori* subband error and noise-free *a posterior* subband error as follows

$$\varepsilon_{i,a}(k) = \mathbf{u}_i^T(k)[\mathbf{w}_o - \mathbf{w}(k)] \tag{5}$$

$$\varepsilon_{i,p}(k) = \mathbf{u}_i^T(k)[\mathbf{w}_o - \mathbf{w}(k+1)]. \tag{6}$$

Then, (2) can be rewritten as

$$e_{i,D}(k) = \varepsilon_{i,a}(k) + \eta_{i,D}(k) \tag{7}$$

where $\eta_{i,D}(k)$ is the *i*th subband noise with zero-mean and variance $\sigma^2_{\eta_{i,D}} = \sigma^2_\eta / N$, which is obtained by partitioning the system noise $\eta(n)$ [10].

Subtracting (4) from $\mathbf{w}_o$ and pre-multiplying $\mathbf{u}_i^T(k)$, we have

$$\begin{aligned}\varepsilon_{i,p}(k) = {} & \varepsilon_{i,a}(k) - \mu_i(k)e_{i,D}(k) \\ & - \sum_{\substack{j=0 \\ i \neq j}}^{N-1} \mu_i(k) \frac{\mathbf{u}_i^T(k)\mathbf{G}(k)\mathbf{u}_j(k)e_{j,D}(k)}{\mathbf{u}_j^T(k)\mathbf{G}(k)\mathbf{u}_j(k)}\end{aligned}. \tag{8}$$

For simplifying (8), we can use the approximation $\mathbf{u}_i^T(k)\mathbf{G}(k)\mathbf{u}_j(k) \approx 0, i \neq j$, due to the commonly used assumption that the subband input signals are almost mutually exclusive [6]-[8] and are independent of the diagonal elements of $\mathbf{G}(k)$ [20]. Thus, combining (7) and (8), we can rearrange (8) as

$$\varepsilon_{i,p}(k) = [1 - \mu_i(k)]\varepsilon_{i,a}(k) - \mu_i(k)\eta_{i,D}(k). \tag{9}$$

Taking the square and mathematical expectation of both sides of (9), we obtain

$$E\left[\varepsilon_{i,p}^2(k)\right] = [1 - \mu_i(k)]^2 E\left[\varepsilon_{i,a}^2(k)\right] + \mu_i^2(k)\sigma^2_{\eta_{i,D}} \tag{10}$$

where $E[\cdot]$ denotes the mathematical expectation, and $\sigma^2_{\eta_{i,D}} \triangleq E\left[\eta_{i,D}^2(k)\right]$. In (10), we also use an assumption that $\eta_{i,D}(k)$ and $\varepsilon_{i,a}(k)$ are mutually independent. Next, we minimize $E\left[\varepsilon_{i,p}^2(k)\right]$ to obtain the individual variable step size $\mu_i(k)$ for each subband. Taking the first-order derivative of (10) with respect to $\mu_i(k)$ leads to

$$\frac{\partial E\left[\varepsilon_{i,p}^2(k)\right]}{\partial \mu_i(k)} = 2\mu_i(k)\left\{E\left[\varepsilon_{i,a}^2(k)\right] + \sigma^2_{\eta_{i,D}}\right\} - 2E\left[\varepsilon_{i,a}^2(k)\right]. \tag{11}$$

Setting (11) to zero, the time-varying step size $\mu_i(k)$ for $i \in [0, N-1]$ is derived as follows,

$$\mu_i(k) = \frac{E\left[\varepsilon_{i,a}^2(k)\right]}{E\left[\varepsilon_{i,a}^2(k)\right] + \sigma^2_{\eta_{i,D}}}. \tag{12}$$

Generally, the statistical value $E\left[\varepsilon_{i,a}^2(k)\right]$ in (12) is approximated by the time average of the square of

$\varepsilon_{i,a}(k)$, i.e.,

$$E\left[\varepsilon_{i,a}^2(k)\right] \triangleq \sigma_{\varepsilon_{i,a}}^2(k) = \theta\sigma_{\varepsilon_{i,a}}^2(k-1) + (1-\theta)\varepsilon_{i,a}^2(k) \qquad (13)$$

where $\theta$ is the forgetting factor which is chosen by $\theta = 1 - N/\kappa M$, $1 \leq \kappa \leq 6$ [8]. Once the noise-free *a priori* subband error $\varepsilon_{i,a}(k)$ is known, an estimate of $E\left[\varepsilon_{i,a}^2(k)\right]$ can be obtained by (13). Subsequently, (12) can be used to compute the step size $\mu_i(k)$. According to the noniterative shrinkage method described in [18], [19], the value of $\varepsilon_{i,a}(k)$ can be recovered from the subband error signal $e_{i,D}(k)$, i.e.,

$$\varepsilon_{i,a}(k) = \text{sgn}\left(e_{i,D}(k)\right)\max\left(\left|e_{i,D}(k)\right| - t_i,\ 0\right) \qquad (14)$$

where sgn(·) denotes the sign function, and the threshold parameter $t_i$ is chosen as $t_i = \sqrt{\lambda \sigma_{\eta_{i,D}}^2}$. Based on extensive simulation results (see Section 4.1), it is found that $\lambda$ in the range 3 to 4 results in good performance.

### 3.2 Discussion

*Remark 1*: As can be seen from (12)-(14), the proposed VSS strategy does not depend on the proportionate matrix $\mathbf{G}(k)$. So, it can be directly applied to any PNSAF-type algorithms (e.g., the PNSAF and IPNSAF) to improve their performance in terms of the convergence rate, tracking capability and steady-state error. Here, the IPNSAF is chosen as the reference, due to its robustness to the unknown system with different sparseness degrees. In this case, the diagonal elements of $\mathbf{G}(k)$, i.e., $g_m(k)$ for $m \in [1,\ M]$ are evaluated as [17]

$$g_m(k) = \frac{1-\alpha}{2M} + (1+\alpha)\frac{|w_m(k)|}{2\|\mathbf{w}(k)\|_1 + \xi} \qquad (15)$$

where $\|\cdot\|_1$ indicates the $l_1$-norm of a vector, $\xi$ is a small positive constant to avoid division by zero, and $-1 \leq \alpha \leq 1$. In practice, good choices for the parameter $\alpha$ are 0 or -0.5 [17]. Hence, the combination between the IPNSAF and proposed VSS scheme yields a new algorithm, called the VSS-IPNSAF which is summarized in Table 1.

It has been shown in Appendix that for ensuring the stability of the PNSAF-type algorithms in the mean-square sense, the range of the step size is $0<\mu<2$. As shown in (12), the value of the individual step size at each iteration lies always in the range of $0<\mu_i(k)<1$. Based on the above reasons, therefore, one can say that the proposed VSS-IPNSAF is mean-square stable.

Table 1 Proposed VSS-IPNSAF algorithm

| | |
|---|---|
| **Initializations** | $\mathbf{w}(0)=\mathbf{0}$, $\varepsilon_{i,a}^2(0)=0$ |
| **Parameters** | $\alpha=0$ or $-0.5$, $\theta=1-N/\kappa M$ with $1\leq\kappa\leq 6$ <br> $\xi$, $\delta$ small positive number <br> $t_i=\sqrt{\lambda\sigma_{\eta_{i,D}}^2}$ with $\lambda=3$ to 4 |
| **Adaptive process** | $e_{i,D}(k)=d_{i,D}(k)-\mathbf{u}_i^T(k)\mathbf{w}(k)$ <br><br> $g_m(k)=\dfrac{1-\alpha}{2M}+(1+\alpha)\dfrac{\|w_m(k)\|}{2\|\mathbf{w}(k)\|_1+\xi}$ <br><br> $\mathbf{w}(k+1)=\mathbf{w}(k)+\sum_{i=0}^{N-1}\mu_i(k)\dfrac{\mathbf{G}(k)\mathbf{u}_i(k)e_{i,D}(k)}{\mathbf{u}_i^T(k)\mathbf{G}(k)\mathbf{u}_i(k)+\delta}$ <br><br> ---Proposed VSS scheme--- <br><br> $\varepsilon_{i,a}(k)=\text{sign}(e_{i,D}(k))\max(|e_{i,D}(k)|-t_i,\,0)$ <br><br> $\sigma_{\varepsilon_{i,a}}^2(k)=\theta\sigma_{\varepsilon_{i,a}}^2(k-1)+(1-\theta)\varepsilon_{i,a}^2(k)$ <br><br> $\mu_i(k)=\dfrac{\sigma_{\varepsilon_{i,a}}^2(k)}{\sigma_{\varepsilon_{i,a}}^2(k)+\sigma_{\eta_{i,D}}^2}$ |

*Remark 2*: In table 2, the computational burden of several adaptive algorithms in terms of the total number of additions, multiplications, divisions, comparisons and absolutions for each fullband input sample and their data memory usage are provided, where the VSS-IPNLMS and VSS-MIPAPA belong to the family of fullband algorithms. Compared with the IPNSAF with the fixed step size, the additional computation of the proposed VSS-IPNSAF stems from (12)-(14) which require 4 multiplications, 3 additions, and 1 comparison for each fullband input sample; and an additional memory of size 5*N* is required for storing intermediate variables to update the step sizes. Fortunately, this slight increase in computational complexity can be compensated by its excellent performance. It is worth noting that the VSS-IPNSAF has almost the same computational complexity as the VSS-IPNLMS especially for a long adaptive filter (i.e., $M \gg L$); while the VSS-MIPAPA requires large computational burden which is at least $P^2$ times than the VSS-IPNLMS since it also requires an additional direct matrix inversion operator of size $P \times P$ except for table 2, even if these two algorithms have better convergence performance than the VSS-IPNLMS for the colored input.

**Table 2** Computational complexity of various adaptive algorithms for each fullband input sample and data memory usage. The integer $L$ denotes the length of the prototype filter of the filter bank, $P$ denotes the affine projection order for the VSS-MIPAPA, and the $UR$ in the SM-IPNSAF denotes the update ratio, where $0 < UR < 1$.

| Algorithms | Multiplications | Additions | Divisions | Comparisons | Absolutions | Data memory usage |
|---|---|---|---|---|---|---|
| NSAF | $3M+3NL+1$ | $3M+3N(L-1)$ | 1 | 0 | 0 | $M(N+1)+3NL+4N+2$ |
| VSSM-NSAF [8] | $6M+3NL+8$ | $5M+3N(L-1)+3$ | 3 | 0 | 0 | $M(2N+1)+3NL+8N+2$ |
| IPNSAF [17] | $5M+3NL+1$ | $3M+3N(L-1)+2M/N$ | $M/N+1$ | 0 | $M/N$ | $M(2N+1)+3NL+4N+3$ |
| SM-IPNSAF [17] | $M+3NL+1+$ $4M \times UR$ | $M+3N(L-1)+$ $(2M/N+2M) \times UR$ | $(M/N+2) \times UR$ | 0 | $(M/N) \times UR$ | $M(2N+1)+3NL+7N+3$ |
| Proposed VSS-IPNSAF | $5M+3NL+5$ | $3M+3N(L-1)+5M/N$ | $M/N+1$ | 1 | $M/N+1$ | $M(2N+1)+3NL+9N+3$ |
| VSS-IPNLMS | $5M+1$ | $5M+3$ | $M$ | 0 | $M+1$ | $4M+8$ |
| VSS-MIPAPA [27] | $(P^2+3)M+P^2+6P$ | $(P^2+P+1)M+P^2+5P$ | $M+P$ | 0 | $M+P$ | $2MP+P^2+3M+2P+3$ |

## 4 Simulation results

In this section, to evaluate the proposed VSS-IPNSAF algorithm, simulations are performed in the context of acoustic echo cancellation. In our simulations, a realistic sparse echo path $\mathbf{w}_o$ with $M = 512$ taps to be estimated is shown in Fig. 2. The colored input signal $u(n)$ at the far-end is either a first-order autoregression, AR(1), process with a pole at 0.95 or a true speech signal. The white Gaussian noise $\eta(n)$ is added to the output-end of the echo path to give a signal-to-noise ratio (SNR) of 30dB or 20dB. The cosine modulated filter bank is used for all SAF algorithms [5], [21]. And, to maintain 60 dB stopband attenuation, the length of the prototype filter is 16, 32, and 64 for number of subbands $N$ = 2, 4, and 8, respectively. It is assumed that the noise variance $\sigma_\eta^2$ is known, because it can be easily estimated online [8], [9], [22]. The normalized mean-squared-deviation (NMSD), $10\log_{10}\left[\|\mathbf{w}_o - \mathbf{w}(k)\|_2^2 / \|\mathbf{w}_o\|_2^2\right]$, is used to measure the algorithm performance, unless otherwise specified. All results are obtained by averaging over 25 independent runs.

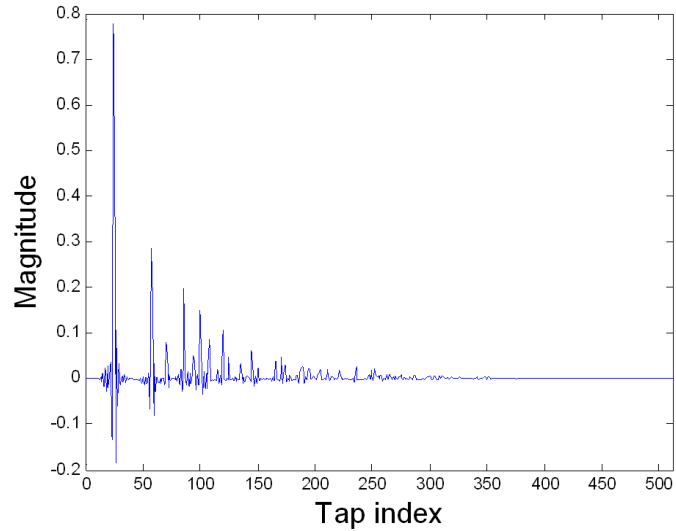

**Fig. 2.** Sparse echo path.

### 4.1 Effect of $\lambda$ and $N$

In this subsection, the AR(1) process is used as the input signal. First, we compare the performance of the VSS-IPNSAF using different number of subbands (i.e., $N$ = 2, 4, and 8), as shown in Fig. 3. As expected, a large number of subbands (e.g., $N$ = 8) can speed up the convergence of the algorithm as compared to a small one (e.g., $N$ = 2). Moreover, this phenomenon will be not obvious when number of subbands is larger than a certain value (in this case is 4). The reason is behind this phenomenon is that each decimated subband input signal is closer to a white signal for a large number of subbands. In other words, the larger number of subbands is, the much stronger decorrelating capability of the VSS-IPNSAF for the colored input signal is. In the following simulations, therefore, number of subbands is chosen as $N$ = 4 for AR(1) input; and we choose $N$ = 8 for speech input since it is nonstationary and highly colored.

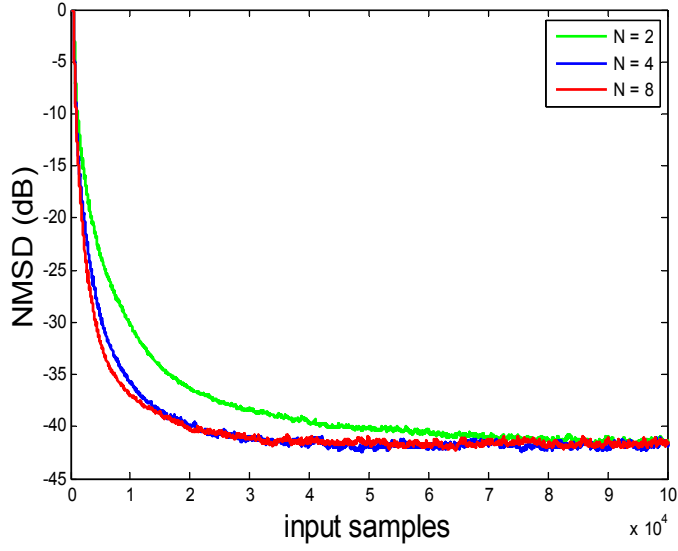

**Fig. 3.** The NMSD curves of the VSS-IPNSAF using different *N* and the fullband VSS-IPNLMS algorithms. Parameters: $\alpha=0$, $\delta=0.001$, $\xi=0.001$ and $\lambda=3$.

Next, Fig. 4 shows the effect of $\lambda$ (which is used for setting the threshold $t_i$ in (14)) on the performance of the VSS-IPNSAF. One can observe from this result that a larger $\lambda$ value (e.g., $\lambda=5$) leads to a lower steady-state error, but slows the convergence rate. A suitable range of $\lambda$ in the VSS-IPNSAF is $3 \leq \lambda \leq 4$ to obtain a good balance between the convergence rate and steady-state error.

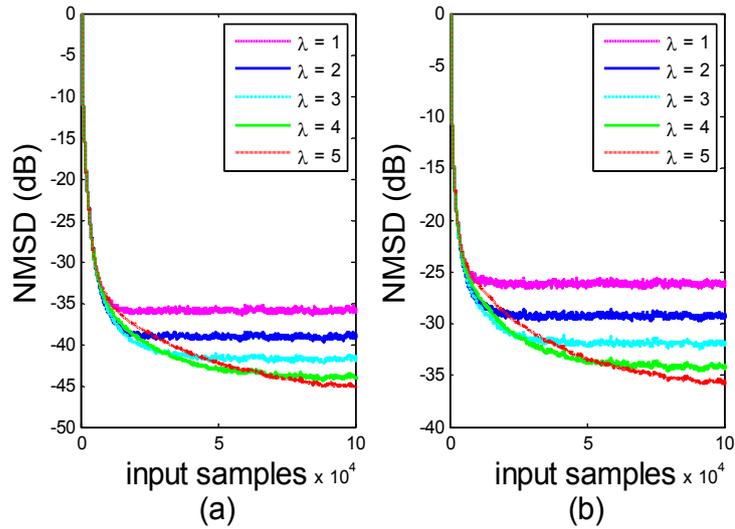

**Fig. 4.** The NMSD curves of the VSS-IPNSAF using different values of $\lambda$. (a) $N=4$, SNR = 30 dB. (b) $N=4$, SNR = 20 dB. Choice of the parameters is the same as Fig. 3.

### 4.2 Comparison of various subband algorithms

In this part, the performance of the VSS-IPNSAF is compared with that of some existing SAF algorithms, i.e., the VSSM-NSAF [8], IPNSAF and SM-IPNSAF [17] algorithms. The SM-IPNSAF is compared because it can be considered a VSS version of the IPNSAF. In this algorithm, the step size is adjusted by

$$\mu_{SM,\ i}(k) = \begin{cases} 1 - \dfrac{\vartheta}{|e_{i,D}(k)|}, & \text{if } |e_{i,D}(k)| > \vartheta \\ 0, & \text{otherwise} \end{cases} \quad (16)$$

where the bound parameter $\vartheta$ is set by $\vartheta = \sqrt{\gamma \sigma_\eta^2 / N}$ [17]. To fairly compare these algorithms, we choose $\alpha = 0$, $\delta = 0.001$ and $\xi = 0.001$ for all IPNSAF-type algorithms, and other parameters are chosen as the recommended values in the literatures [8, 17]. Also, to assess the tracking performances of these algorithms, the echo path is multiplied by a factor of $-1$ at input sample index 140, 000 for AR(1) input or 200, 000 for speech input.

*1) AR(1) input*

Fig. 5 shows the NMSD results of these algorithms for AR(1) input. As can be seen, the fixed step size controls the convergence, tracking and steady-state properties of the original IPNSAF. That is, a large step-size yields fast convergence rate and good tracking capability, but increases the steady-state error, and vice versa. Both the SM-PNSAF and proposed VSS-IPNSAF can effectively address this problem of the IPNSAF, since they both use the time-varying step sizes, see (12) and (16). However, the VSS-IPNSAF achieves a decrease (about 7dB) in the steady-state error as compared to the SM-IPNSAF. In addition, the VSSM-NSAF has much slower convergence rate for sparse echo path than the VSS-IPNSAF, because it is a non-proportionate VSS algorithm.

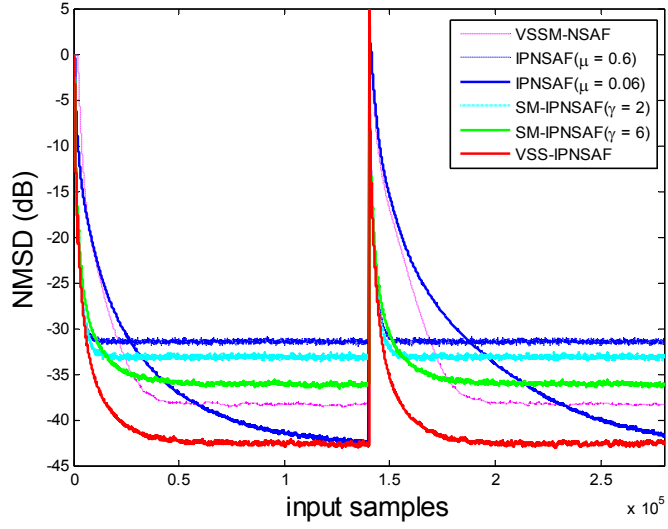

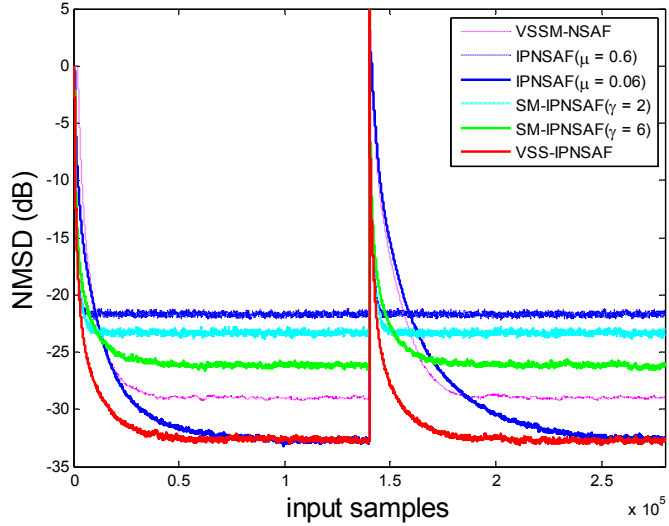

**Fig. 5.** The NMSD curves of various SAF algorithms for AR(1) input. (a) $N = 4$, SNR = 30 dB. (b) $N = 4$, SNR = 20 dB. VSSM-NSAF: $\kappa = 6$ [8]; VSS-IPNSAF: $\kappa = 1$ and $\lambda = 3.5$.

*2) Speech input*

Fig. 6 shows the NMSD results of all SAF algorithms using a true speech as input in the case of 30dB and/or 20dB. It is clear that the proposed VSS-IPNSAF outperforms other SAF algorithms (i.e., the VSSM-NSAF, IPNSAF, and SM-IPNSAF) in terms of the convergence rate and steady-state NMSD. Fig. 7 compares the echo return loss enhancement (ERLE) performance of these algorithms, where the simulation condition is the same as Fig. 6. The ERLE is defined as [25], [26]

$$\mathrm{ERLE}(n) = 10\log 10\left\{E\left[d^2(n)\right]\Big/E\left[e^2(n)\right]\right\}, \text{ (dB)} \tag{17}$$

As one can see, even if the ERLE is used as performance criterion, the VSS-IPNSAF still has better performance than the VSSM-NSAF, IPNSAF, and SM-IPNSAF algorithms in terms of the convergence rate and steady-state ERLE.

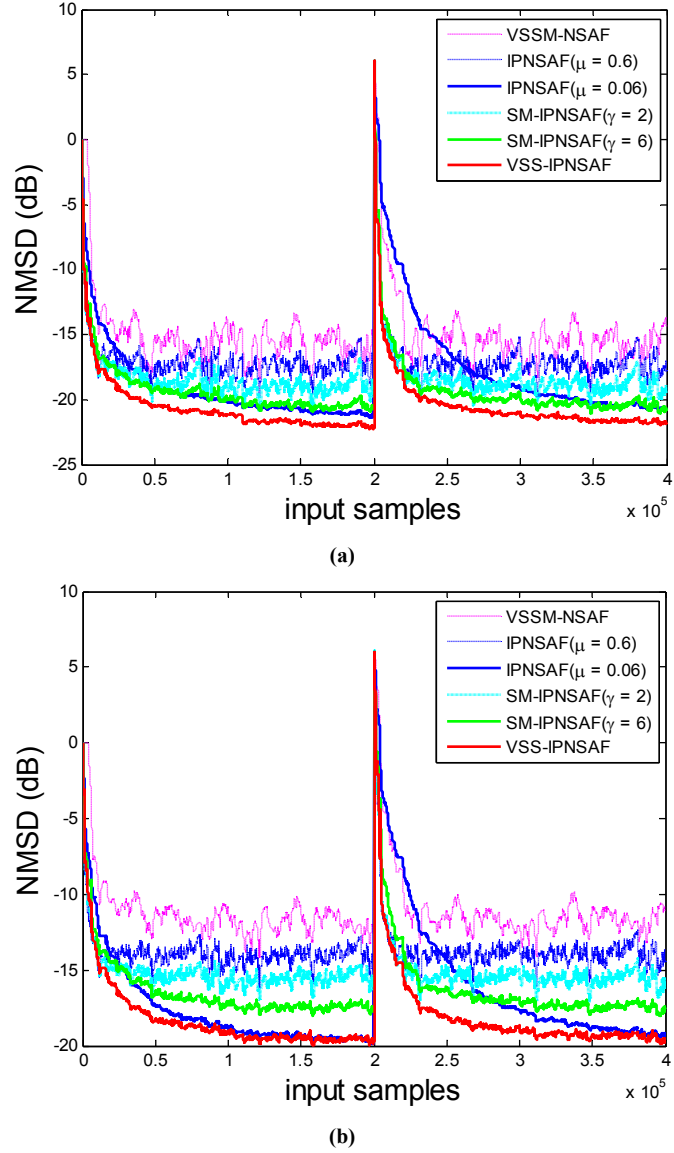

**Fig. 6.** The NMSD curves of various SAF algorithms for speech input. (a) $N = 8$, SNR = 30 dB. (b) $N = 8$, SNR = 20 dB. The choice of other parameters is the same as Fig. 5.

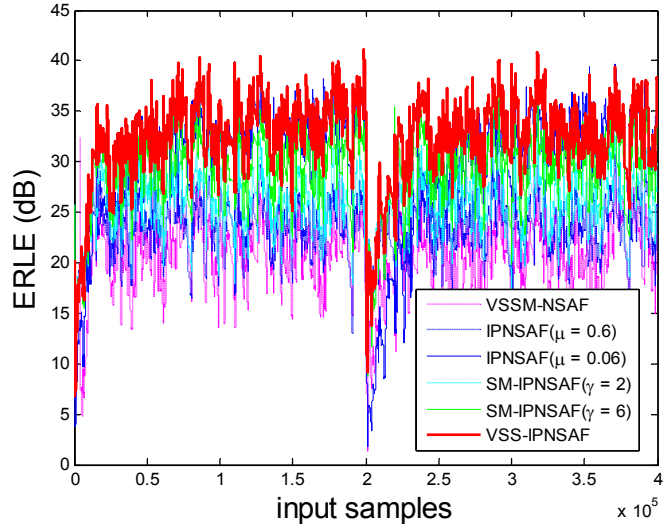

(a)

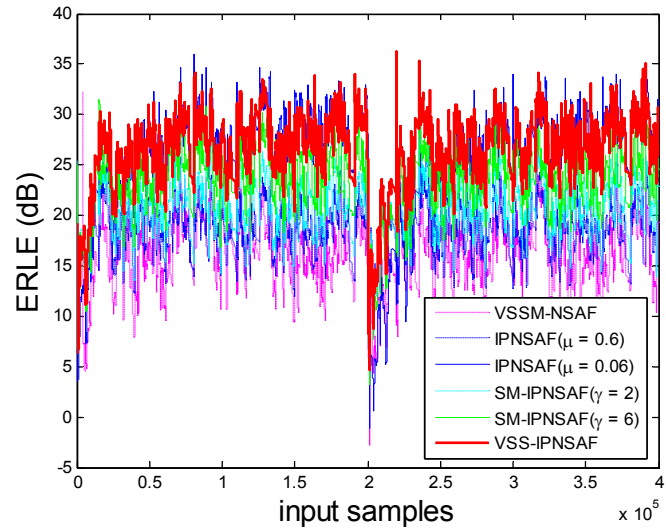

(b)

**Fig. 7.** The ERLE curves of various SAF algorithms for speech input. (a) $N = 8$, SNR = 30 dB. (b) $N = 8$, SNR = 20 dB.

### 4.3 Comparison with fullband algorithms

In this example, the performance of the VSS-IPNSAF algorithm is compared with that of two fullband VSS algorithms (i.e., the VSS-IPNLMS algorithm and the VSS-MIPAPA algorithm presented in [27]), as shown in Fig. 8. To obtain a fair assessment, these three algorithms employ the same formula, i.e., (15) for the calculation of the proportionate matrix $\mathbf{G}(k)$, and thus the corresponding proportionate parameters are set to $\alpha = 0$ and $\xi = 0.001$.

And, the VSS-IPNLMS has the similar VSS scheme as the VSS-IPNSAF. The affine projection order for the VSS-MIPAPA is chosen as $P=4$. As can be seen, both the VSS-MIPAPA and proposed VSS-IPNSAF have faster convergence rate than the VSS-IPNLMS for the colored input signal, owing mainly to their inherent decorrelating properties in the subband domain and the time domain, respectively. However, the VSS-MIPAPA requires much larger computational cost than the VSS-IPNSAF, see table 2 for details.

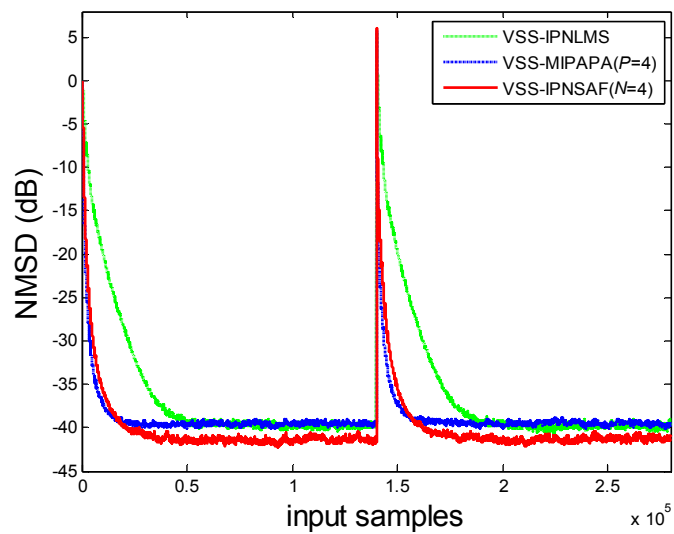

(a)

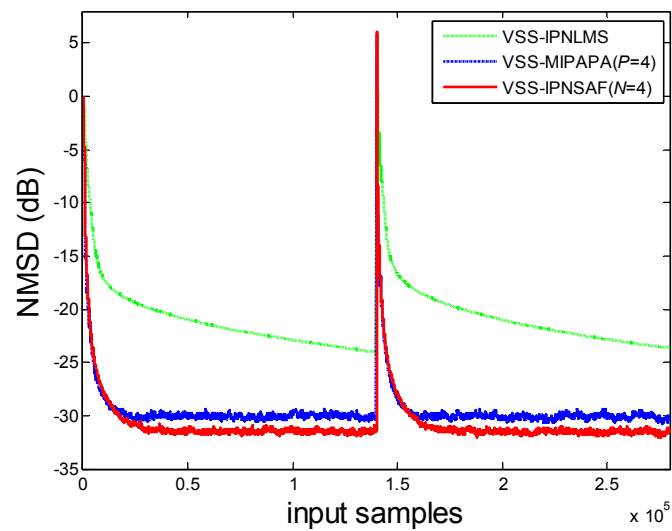

(b)

**Fig. 8.** The NMSD curves of the VSS-IPNSAF and two fullband algorithms for AR(1) input. (a) SNR = 30 dB. (b) SNR = 20 dB. VSS-IPNLMS: $\kappa=1$ and $\lambda=3.5$; VSS-MIPAPA: $\varepsilon=0.001$ and $K=2$ [27]; VSS-IPNSAF: $\kappa=1$ and $\lambda=3.5$.

## 5 Conclusion

In this study, we derived a VSS version of the IPNSAF from the minimization of the mean-square noise-free *a posterior* subband errors. And, in order to recover the noise-free *a priori* subband error signal from the noisy subband error signal, a noniterative shrinkage method was employed. Interestingly, the proposed VSS scheme is easily applicable to any other PNSAF algorithms to further improve the performance, since it is independent of the computing method of proportionate matrix. Simulation results for acoustic echo cancellation have verified that the proposed VSS method is effective.

**Acknowledgment**

This work was partially supported by National Science Foundation of P.R. China (Grant: 61271340, 61571374 and 61433011), and the Fundamental Research Funds for the Central Universities (Grant: SWJTU12CX026).

The authors would like to thank Prof. C. Paleologu at University Politehnica of Bucharest, Bucharest, Romania, for supplying the true echo path and speech signal used in our simulations.

**Appendix**

In this section, we analyze the convergence condition of PNSAF-type algorithms in the mean-square sense based on the energy conservation theory [23]. To the best of our knowledge, it has not been reported in the previous literatures.

Rewrite (3) in matrix form as

$$\mathbf{w}(k+1) = \mathbf{w}(k) + \mu \mathbf{G}(k)\mathbf{U}(k)\mathbf{\Lambda}^{-1}(k)\mathbf{e}_D(k) \tag{17}$$

where

$$\mathbf{U}(k) = [\mathbf{u}_0(k),\ \mathbf{u}_1(k),\ ...,\ \mathbf{u}_{N-1}(k)] \tag{18}$$

$$\mathbf{e}_D(k) = [e_{0,D}(k),\ e_{1,D}(k),\ ...,\ e_{N-1,D}(k)]^T \tag{19}$$

$$\mathbf{\Lambda}(k) = \mathrm{diag}\{\mathbf{u}_0^T(k)\mathbf{G}(k)\mathbf{u}_0(k),\ \mathbf{u}_1^T(k)\mathbf{G}(k)\mathbf{u}_1(k),\ ...,\ \mathbf{u}_{N-1}^T(k)\mathbf{G}(k)\mathbf{u}_{N-1}(k)\}. \tag{20}$$

Subtracting $\mathbf{w}_o$ from both sides of (17), we have

$$\widetilde{\mathbf{w}}(k+1) = \widetilde{\mathbf{w}}(k) + \mu \mathbf{G}(k)\mathbf{U}(k)\mathbf{\Lambda}^{-1}(k)\mathbf{e}_D(k) \tag{21}$$

where $\widetilde{\mathbf{w}}(k) = \mathbf{w}(k) - \mathbf{w}_o$ is the tap-weight error vector.

Let us define the noise-free *a priori* subband error vector and noise-free *a posterior* vector as

$$\boldsymbol{\varepsilon}_a(k) = \mathbf{U}^T(k)\widetilde{\mathbf{w}}(k) \triangleq [\varepsilon_{0,a}(k),\ \varepsilon_{1,a}(k),\ ...,\ \varepsilon_{N-1,a}(k)]^T, \tag{22}$$

$$\boldsymbol{\varepsilon}_p(k) = \mathbf{U}^T(k)\widetilde{\mathbf{w}}(k+1) \triangleq [\varepsilon_{0,p}(k),\ \varepsilon_{1,p}(k),\ ...,\ \varepsilon_{N-1,p}(k)]^T. \tag{23}$$

Combining (21)-(23), we obtain

$$\boldsymbol{\varepsilon}_a(k) - \boldsymbol{\varepsilon}_p(k) = \mu \mathbf{U}^T(k)\mathbf{G}(k)\mathbf{U}(k)\boldsymbol{\Lambda}^{-1}(k)\mathbf{e}_D(k). \tag{24}$$

Assuming that $\mathbf{M}(k) = \mathbf{U}^T(k)\mathbf{G}(k)\mathbf{U}(k)$ is invertible, and substituting (24) into (21), we get

$$\widetilde{\mathbf{w}}(k+1) = \widetilde{\mathbf{w}}(k) + \mathbf{G}(k)\mathbf{U}(k)\mathbf{M}^{-1}(k)\left[\boldsymbol{\varepsilon}_p(k) - \boldsymbol{\varepsilon}_a(k)\right]. \tag{25}$$

By taking the squared Euclidean norm of both sides of (25), we find that the following relation should hold:

$$\left\|\widetilde{\mathbf{w}}(k+1)\right\|^2 + \boldsymbol{\varepsilon}_a^T(k)\boldsymbol{\Gamma}(k)\boldsymbol{\varepsilon}_a(k) = \left\|\widetilde{\mathbf{w}}(k)\right\|^2 + \boldsymbol{\varepsilon}_p^T(k)\boldsymbol{\Gamma}(k)\boldsymbol{\varepsilon}_p(k) \tag{26}$$

where $\|\cdot\|$ denotes the Euclidean norm, and $\boldsymbol{\Gamma}(k) = \left[\mathbf{M}^{-1}(k)\right]^T \mathbf{U}^T(k)\mathbf{G}(k)\mathbf{G}(k)\mathbf{U}(k)\mathbf{M}^{-1}(k)$. It is worthy to note that (26) is an exact energy relation, since no assumptions and/or approximations are used to establish it. It shows how the energies of the tap-weight error vectors at two adjacent time instants are related to the energies of the *a priori* and *a posteriori* error vectors.

Taking the expectation on both sides of (26), and rearrange it as

$$\text{MSD}(k+1) - \text{MSD}(k) = E\left[\boldsymbol{\varepsilon}_p^T(k)\boldsymbol{\Gamma}(k)\boldsymbol{\varepsilon}_p(k)\right] - E\left[\boldsymbol{\varepsilon}_a^T(k)\boldsymbol{\Gamma}(k)\boldsymbol{\varepsilon}_a(k)\right] \tag{27}$$

where $\text{MSD}(k) \triangleq E\left[\left\|\widetilde{\mathbf{w}}(k)\right\|^2\right]$ is the mean squared deviation (MSD).

Substituting (24) into the first term of the right-hand side of (27), we obtain

$$\text{MSD}(k+1) - \text{MSD}(k) = \mu^2 E\left[\left(\mathbf{M}(k)\boldsymbol{\Lambda}^{-1}(k)\mathbf{e}_D(k)\right)^T \boldsymbol{\Gamma}(k)\left(\mathbf{M}(k)\boldsymbol{\Lambda}^{-1}(k)\mathbf{e}_D(k)\right)\right] \\ -2\mu E\left[\boldsymbol{\varepsilon}_a^T(k)\boldsymbol{\Gamma}(k)\mathbf{M}(k)\boldsymbol{\Lambda}^{-1}(k)\mathbf{e}_D(k)\right] \tag{28}$$

We again use the approximation $\mathbf{u}_i^T(k)\mathbf{G}(k)\mathbf{u}_j(k) \approx 0$, $i \neq j$, i.e., the off-diagonal elements of the matrix $\mathbf{M}(k)$ are negligible, then (28) can be simplified as

$$\text{MSD}(k+1) - \text{MSD}(k) = \mu^2 E\left[\mathbf{e}_D^T(k)\boldsymbol{\Gamma}(k)\mathbf{e}_D(k)\right] - 2\mu E\left[\boldsymbol{\varepsilon}_a^T(k)\boldsymbol{\Gamma}(k)\mathbf{e}_D(k)\right] \tag{29}$$

For ensuring the stability of the algorithm, the MSD must decrease iteratively, i.e., $\text{MSD}(k+1) - \text{MSD}(k) < 0$.

Thus, the step size $\mu$ has to satisfy the inequality

$$0 < \mu < 2 \frac{E\left[\boldsymbol{\varepsilon}_a^T(k)\boldsymbol{\Gamma}(k)\mathbf{e}_D(k)\right]}{E\left[\mathbf{e}_D^T(k)\boldsymbol{\Gamma}(k)\mathbf{e}_D(k)\right]} \quad (30)$$

If we consider the ideal situation that the system noise $\eta(n)$ in Fig. 1 is negligible, the noise-free *a priori* subband error vector $\boldsymbol{\varepsilon}_a(k)$ is equal to the decimated subband error vector $\mathbf{e}_D(k)$. Therefore, in the absence of the system noise, the convergence condition of PNSAF-type algorithms in the mean-square sense is that the step size is in the range of

$$0 < \mu < 2. \quad (31)$$